\documentclass{article}
\def\DH{\rm I\kern-1.5pt\rm H\kern-1.5pt\rm I}

\input epsf

\def\DR{\rm I\kern-1.45pt\rm R}
\def\DC{\kern2pt {\hbox{\sqi I}}\kern-4.2pt\rm C}

\newcommand{\ba}{\begin{array}}
\newcommand{\ea}{\end{array}}
\newcommand{\be}{\begin{equation}}
\newcommand{\ee}{\end{equation}}
\newcommand{\bea}{\begin{eqnarray}}
\newcommand{\eea}{\end{eqnarray}}
\newcommand{\bi}{\begin{itemize}}
\newcommand{\ei}{\end{itemize}}

\begin{document}
\thispagestyle{empty}
\begin{center}
{\bf \Large  Second Hopf map and Yang-Coulomb  system on $5d$ (pseudo)sphere}\\
\vspace{0.5 cm} {\large Stefano Bellucci$^1$, Francesco
Toppan$^{2}$  and Vahagn Yeghikyan$^{3}$}
\end{center}
{\it ${}^1$INFN-Laboratori Nazionali di Frascati, Via E. Fermi 40,
00044 Frascati, Italy}\\
$\;^2${\it CBPF, Rua Dr. Xavier Sigaud 150, cep 22290-180, Rio de Janeiro (RJ), Brazil}\\
$\;^3${\sl  Yerevan State University, A. Manoogian St., 1,  Yerevan,
0025, Armenia}
\begin{abstract}
Using the second Hopf map, we perform the reduction of the
eight-dimensional (pseudo)spherical  (Higgs)oscillator to a
five-dimensional system interacting with a Yang monopole. Then,
using a standard trick, we obtain, from the latter system,  the
pseudospherical and spherical generalizations of the Yang-Coulomb
system (the five dimensional analog of MICZ-Kepler system). We
present the whole set of its constants of motions, including the
hidden symmetry generators given by the analog of Runge-Lenz
vector. In the same way, starting from the eight-dimensional
anisotropic inharmonic Higgs oscillator, we construct the
integrable (pseudo)spherical generalization of the Yang-Coulomb
system with the Stark term.
\end{abstract}
\section{Introduction}
It is well-known that in some cases the hidden symmetries of the
oscillator and the Coulomb system can be related. More precisely,  both in classical and quantum cases,
the transformation $r=R^2$ converts
 the $(p+1)-$dimensional radial Coulomb problem
to a  $2p-$dimensional
 radial oscillator  ($r$ and $R$
denote the radial coordinates of Coulomb and oscillator systems,
 respectively). The angular parts of these systems, which are $(2p-1)$- and $p$-dimensional spheres,
 can  be related  with each other in the distinguished cases
 $p=1,2,4,8$, due to the existence of the Hopf maps $S^{2p-1}/S^p=
 S^{p-1}$  for $p=1,2,4,8$ (see the review  \cite{ta} and refs therein). Hence, in these cases, one can establish a
complete  correspondence
 between the Coulomb and the oscillator systems. For the first three cases this correspondence has been established many years ago.
The corresponding transformations are known in literature as
Bohlin (or Levi-Civita) \cite{bohlin}, Kustaanheimo-Stiefel
\cite{ks} and Hurwitz \cite{h} transformations.
These transformations  imply
 the reduction of the oscillator system by an  action of the group $Z_2$,
$U(1)$, $SU(2)$ and yield, indeed, to systems which are more general than the Coulomb one:
these systems inherit the Coulomb symmetry but are specified by the presence of
a monopole. In the case of the $Z_2$ group ($p=1$) it is the two-dimensional Coulomb problem with spin $1/2$
anyon (magnetic flux) \cite{ntt}; in the case of the $U(1)$ group ($p=2$) it is the so-called MICZ-Kepler system,
the generalization of the three-dimensional Coulomb system in presence of a Dirac monopole \cite{micz}; in the case of the
$SU(2)$ group ($p=4$) it is the so-called Yang-Coulomb (or $SU(2)$-Kepler) system, the generalization of the five-dimensional
Coulomb system in presence of a Yang monopole
\cite{su2}.

On the other hand, the  oscillator and Coulomb systems admit
generalizations to a $d-$di\-men\-si\-onal sphere and a two-sheet
hyperboloid (pseudosphere) with a radius $R_0$
 given by the potentials \cite{sphere,sphere1}
\begin{equation}
V_{osc}=\frac{\omega^2 }{2}\frac{{\bf x}^2}{{x}^2_{d+1}},\quad
V_{C}= -\frac{\gamma}{}\frac{x_{d+1}}{|{\bf x}|},
\label{v}\end{equation} where ${\bf x}, x_{d+1}$ are the
(pseudo)Euclidean coordinates of the ambient space
$\DR^{d+1}$($\DR^{d.1}$): $\epsilon{\bf x}^2+ x^2_{d+1}=R_0^2$,
$\epsilon=\pm 1$.
Here $\epsilon=+1$ corresponds to the sphere and
$\epsilon=-1$ corresponds to  the pseudosphere. These systems
possess nonlinear hidden symmetries providing them with properties
similar to those
 of the conventional oscillator and Coulomb systems. Detailed considerations of these systems can be found in
 \cite{garb} and refs therein.
 The relation between these systems has been established in \cite{pogos} for the cases $p=1,2$ only,
  though it was clear from the considerations there, that it could be straightforwardly extended to the higher dimensional case $p=4$.
  Let us notice that both the oscillators on the sphere and pseudosphere result, upon the mentioned reduction, in Coulomb-like
  systems on the {\sl pseudosphere}. For example, the $p=2$ case yields a system with Coulomb symmetry
   which can be identified with the pseudospherical MICZ-Kepler system. Then, after an obvious ``Wick rotation" it can be mapped
   in the spherical MICZ-Kepler system suggested  in \cite{kurochkin}.
   Moreover, in this way, starting from the appropriate anisotropic inharmonic four-dimensional (pseudo)spherical Higgs oscillator,
   one can obtain the integrable (pseudo)spherical generalization with the Stark term \cite{anosc}.
   Hence, the extension of the (pseudo)spherical oscillator-Coulomb correspondence to the $p=4$ case is
   an important task: it should give us the integrable  generalizations of the five-dimensional Yang-Coulomb system,
   including the systems with Stark term.
    The solution of this task is  the purpose of our paper.
We shall use the procedure of the Lagrangian $SU(2)$-reduction of eight-dimensional system considered in our recent paper
\cite{2nd} in the context of supersymmetric mechanics. We will restrict ourself to classical considerations only.
The (pseudo)spherical Yang-Coulomb system, obtained in the present paper,
besides the presence of a Yang monopole, is specified by the presence of a specific centrifugal term
\be
\Delta U =\frac{s^2}{2g(r)r^2},
\label{delta}\ee
where $g(r)dx^\mu dx^\mu$ is the conformal invariant metric on the (pseudo)sphere and $s^2$ is the square of the isospin
 of the system. Upon quantization it should be replaced by $\hbar^2s(s+1)$, $s=0,\pm 1/2,\pm 1,\ldots$.
In \cite{mny} it was demonstrated that the five-dimensional rotationally invariant
system of the  Yang monopole, supplied by the addition of the above potential, preserves the analytic
 form of the energy spectrum. The only change in the spectrum of the system is the range
 of validity of the orbital quantum number. Its lower bound shifts from zero to $|s|$.
Hence, one  can immediately write down the spectrum of the (pseudo)spherical Yang-Coulomb system,
taking into account general statements. But why we are so sure that the
quantum mechanical reduction of the eight-dimensional Higgs oscillator to the (pseudo)spherical Yang-Coulomb system
should lead to a system with such spectrum: in general, reduction and quantization are not commuting procedures.
We refer to the second reference in \cite{pogos}, where the consistency of the quantum mechanical spectrum of the
eight-dimensional Higgs oscillator and pseudospherical Yang-Coulomb system has been demonstrated.
Hence, for the unperturbed quantum-mechanical systems everything works finely.
With the quantum mechanics of the (pseudo)spherical Yang-Coulomb systems with Stark term the situation is more complicated.
In contrast with the spherically symmetric systems, the impact of the monopole  is less trivial.
Even in the three-dimensional planar case, the presence of a (Dirac) monopole
 essentially changes the spectrum of the MICZ-Kepler system with Stark term (and of its modification)
 obtained within perturbation theory\cite{stark}. Similar calculations for the (pseudo)spherical case yield to technical
complications, which we hope to address in future studies.

The paper is arranged as follows:

In the Second Section we present the explicit description of the
 second Hopf map in terms  needed for our purposes and employ it
 to reduce the
eight-dimensional bosonic  system to a lower dimensional system with
SU(2) monopole.

In the Third Section we apply the previous construction to the oscillator on eight-dimensional (pseudo)sphere
and get, from the reduced system, the  five-dimensional (pseudo)spherical generalization of Yang-Coulomb system.
In a similar way we get the five-dimensional (pseudo)spherical generalization of Yang-Coulomb system
 with Stark term.
\setcounter{equation}{0}

\section{$SU(2)$ reduction and second Hopf map}
For the description of the second Hopf map $S^7/S^3=S^4$ we first
introduce five  $8 \times 8$ matrices $\Gamma^\mu$ \be
\left\{\Gamma^\mu,\Gamma^\nu\right\}=2\delta^{\mu\nu}{\bf 1}_8 \ee
 with the following relations: \be
\Gamma^1=\tau_A\otimes\tau_1\otimes\tau_A,\quad
\Gamma^2=\tau_A\otimes\tau_2\otimes\tau_A,\quad
\Gamma^3=\tau_A\otimes\tau_A\otimes{\bf 1}_2,\quad
\Gamma^4=\tau_1\otimes{\bf 1}_2\otimes{\bf 1}_2,\quad
\Gamma^5=\tau_2\otimes{\bf 1}_2\otimes{\bf 1}_2, \ee where \be
\tau_1= \left( {\begin{array}{*{20}c}
   0 & 1  \\
   1 & 0  \\
\end{array}} \right),\quad
 \tau_2= \left( {\begin{array}{*{20}c}
   1 & 0  \\
   0 & -1  \\
\end{array}} \right),\quad
\tau_A= \left( {\begin{array}{*{20}c}
   0 & 1  \\
   -1 & 0  \\
\end{array}} \right)
\quad {\bf 1 }_2= \left( {\begin{array}{*{20}c}
   1 & 0  \\
   0 & 1  \\
\end{array}} \right),
\label{tau}\ee where $\left\{A,B\right\}$ denotes the anticommutator.
For our purposes we have also to introduce three $8\times 8 $
antisymmetric matrices $\Sigma_a$: \be \Sigma^1 =\frac{1}{2}{\bf
1}_2\otimes\tau_A\otimes\tau_1,\quad \Sigma^2 =\frac{1}{2}{\bf
1}_2\otimes\tau_A\otimes\tau_2,\quad \Sigma^3 =\frac{1}{2}{\bf
1}_2\otimes\bf{1}_2\otimes\tau_A. \label{sigma}\ee which commute
with all matrices $\Gamma_\mu$, anticommute with each other and
satisfy the $ su(2)$ algebra relations:
\be\left[\Gamma^\mu,\Sigma^i\right]=0,\quad\left\{\Sigma^i,\Sigma^j\right\}=-2\delta^{ij}{\bf
1}_8,\quad
\left[\Sigma^i,\Sigma^j\right]=\varepsilon_{ijk}\Sigma^k\ee
Let us now have an 8 dimensional  conformal flat space with metric g,
parametrized by 8 coordinates $u_A$. We also
consider 5 functions $x_\mu$, which are connected with $u_i$ by the
following relations \be x^\mu=u\Gamma^\mu u,\label{brel} \ee where
$u$ is an 8 dimensional column vector with elements $u_A$.

One can can notice that the transformation \be u\rightarrow
\left(\lambda_0 {\bf 1}_8+\lambda_i\Sigma_i\right) u,\quad
\lambda_0^2+\sum{\lambda_i^2}=1\label{trans} \ee leaves invariant
the $x_\mu$ quantities. Therefore, the fibration \ref{brel} identifies
all points which differ by the transformation \ref{trans}. It can be
checked explicitly that \be x_\mu x_\mu\equiv r^2=(u_Au_A)^2\equiv
R^4\label{sq}.\ee Thus, defining the seven- dimensional sphere in
$\DR^{8}$ of radius $R$: ${\bf u}_\alpha{\bf \bar u}_\alpha
=R^2$, we get a $4$-dimensional sphere in $\DR^{5}$ with
radius $r=R^2$, i.e. we obtain the second Hopf map.  Taking into
account the relation \ref{sq} and the fact that the second
relation in \ref{trans} defines the $S^3$ sphere, one can conclude that
the second Hopf map is a fibration of the sphere $S^7$ over $S^3$: $$
S^7/S^3=S^4$$

In order to invert expressions let us introduce
3 additional coordinates: \be z=\frac{u_7-{\bf
i}u_8}{u_5-{\bf i}u_6},\quad \bar z=\frac{u_7+{\bf i}u_8}{u_5+{\bf
i}u_6},\quad \gamma=\arctan{\frac{u_5}{u_6}}  \ee It is easy to
see that the coordinates $z,\bar z,\gamma$ parametrize the sphere
$x_\mu=const$.

The matrices $\Sigma^i$ define a set of vector-fields on $S^3$ that
form the $su(2)$ algebra: \be {\bf
U}_i=u_A\Sigma^i_{AB}\frac{\partial}{\partial u_B}.\ee In terms of
the new coordinates these vector-fields can be written as follows: \be
 {\bf
U}_3= -\frac{1}{2}\frac{\partial}{\partial\gamma} \qquad
{\bf{U}}_2+{\bf i}{\bf{U}}_1={\bf{U}}_+= \frac{{\rm e}^{-2
\imath\gamma}}{4}\left(\left(1+z \bar
z\right)\frac{\partial}{\partial \bar z}+\frac{\imath z}{2}
\frac{\partial}{\partial \gamma}\right), \qquad {\bf{U}}_-=
\overline{{\bf{U}}}_+: \label{vec_2}\ee The one-forms dual to this
set of vector-fields look as follows: \be {\widetilde\Lambda}_3=
2d\gamma +{\bf i}\frac{zd\bar z-\bar z dz}{1+z\bar z} \qquad
{\widetilde\Lambda}_+={\widetilde\Lambda}_2+{\bf
i}{\widetilde\Lambda}_1= 2\frac{{\rm e}^{2\imath \gamma}d\bar
z}{1+z\bar z}\label{13_1}\ee\be {\widetilde\Lambda}_3({\bf
U}_3)={\widetilde\Lambda}_\pm({{\bf U}}_\pm )=1,\qquad
{\widetilde\Lambda}_{\pm}({{\bf U}}_\mp )={\widetilde\Lambda}_\pm
({{\bf U}}_3)= {\widetilde\Lambda}_3({{\bf U}}_\pm)=0 \ee
 We shall
need also another $SU(2)$ group elements parameterizing the sphere
$S^3$ and commuting with (\ref{vec_2}): \be {\bf V}_3=
\frac{1}{2}\frac{\partial}{\partial\gamma}+\imath
\left(z\frac{\partial}{\partial z} -
\bar{z}\frac{\partial}{\partial \bar{z}}\right), ,\qquad {\bf
V}_+=\frac{1}{2}\left(\frac{\partial}{\partial \bar z}
+{z}^2\frac{\partial}{\partial z}
-\imath\frac{z}{2}\frac{\partial}{\partial\gamma}\right) ,\qquad
{\bf V}_-= \overline{{\bf V}}_+: \label{vec}\ee The one-forms dual
to these vector-fields look as follows: \be
\Lambda_3=2h_{3}d\gamma +\imath\frac{\bar{z} d z-z
d\bar{z}}{1+z\bar{z}} \qquad \Lambda_+=\Lambda_2+{\bf
{i}}\Lambda_1=2\imath{\bf h}_+ d\gamma +2\frac{d \bar
z}{1+z\bar{z}}
 \label{13},\ee

\be \Lambda_3({\bf V}_3)=\Lambda_\pm({\bf V}_\pm )=1,\qquad
\Lambda_{\pm}({\bf V}_\mp )=\Lambda_\pm ({\bf V}_3)=
\Lambda_3({\bf V}_\pm)=0, \ee here $h_3, {\bf h_\pm}$ are the
Cartesian coordinates of the ambient space $\DR^3$, defining the
Killing potentials of $S^2$ \be {\bf h}_+=\frac{2z}{1+\bar z z},
\quad { h}_{p+1}=\frac{1-\bar z z}{1+\bar z z}. \label{kil}\ee

 It can be checked directly that the vector fields $U$ and $V$
 commute with each other. Precisely, this pair forms the  the
$so(4)=s(3)\times so(3)$ algebra of isometries of the $S^3$
sphere.
 \be
[{\bf V}_i,{\bf V}_j]=\varepsilon_{ijk}{\bf V}_k,\qquad [{\bf
U}_i,{\bf U}_j]=\varepsilon_{ijk}{\bf U}_k,\qquad [{\bf V}_i,{\bf
U}_j]=0. \label{comm} \ee

Let us consider now the particle on the   $8$-dimensional space
equipped  with the $SU(2)$-invariant  conformal flat metric moving
in a specific potential that depends only on the coordinates $x_\mu$.
In the new parametrization the Lagrangian of that particle has
the following form: \be {\cal L}_{8}=g(x)\dot{u}_A\dot{u}_A +
U(x) = g(x){\dot r}_\mu{\dot r}_\mu + \frac{g
r}{2}\Lambda^{.}_{i}A_i
-\frac{gr}{4}\Lambda^{.}_{i}\Lambda^{.}_{i} + U(x),
 \label{lag0}\ee where
$$ r_\kappa=\frac{x_\kappa}{\sqrt{2(r+x_5)}},\quad \kappa \leq 4,\quad and \quad r_5=\sqrt{\frac{r+x_5}{2}} $$

  Here and
further $\Lambda^{.}_i$ is defined by (\ref{13}), where the
differentials are replaced by the respective time derivatives,
while

\be
A_i=A^i_\beta\dot{x}_\beta=\frac{\eta^i_{\alpha\beta}x_\alpha\dot{x}_\beta}{r\left(r+x_5\right)},
\qquad\eta^i_{\alpha\beta}=\delta_{i\alpha}\delta_{4\beta}-\delta_{4\alpha}\delta_{i\beta}+\varepsilon_{4i\alpha\beta},\quad
i=1,2,3,\quad \alpha,\beta=1,2,3,4\ee

 $\eta^i_{bc}$ is t'Hooft symbol.

It can be seen that  $A_a$ defines the potential of the
the $SU(2)$ Yang
monopole.

  By the use of the Noether constants of motion we can decrease the dimensionality of the system. Due to the  non-Abelian nature of $SU(2)$
group the system will be reduced to a $(5+1)$-dimensional one.

For the correct reduction procedure we have to replace the initial Lagrangian by a variationally equivalent one,
extending the initial configuration space by the new variables $\pi$, $\bar\pi$, $p_\gamma$ which play the role of
momenta conjugated to $z$, $\bar z$, $\gamma$. In other words, we will  replace the sphere $S^3$
(parameterized by $z$, $\bar z$, $\gamma$), by its cotangent bundle $T^*S^3$ parameterized by
coordinate $z$, $\bar z$, $\gamma$, $\pi$, $\bar\pi$, $p_\gamma$.
 Let us  define, on $T^*S^3$,
the Poisson brackets given by the relations \be \{\pi,
z\}=1,\qquad\{\bar\pi, \bar z\}=1,\qquad\{p_\gamma, \gamma\}=1.
\label{pb}\ee We introduce the  Hamiltonian generators $P_a$
corresponding to  the vector fields (\ref{vec}) (replacing  the
derivatives in vector fields ${\bf V}_a$  by corresponding
momenta) \be
 P_+=\frac{P_2-\imath
P_1}{2}=\frac{1}{2}\left(\pi+\bar{z}^2 \bar\pi-\frac{\imath
\bar{z}}{2} p_{\gamma }\right),\quad P_-=\bar P_-,\quad
P_3=\frac{p_\gamma}{2}+\imath\left(z\pi-\bar
z\bar\pi\right).\label{ier}\ee In the same way we introduce the
Hamiltonian generators $I_a$ corresponding to the vector fields
(\ref{vec_2}): \be I_3=-\frac{p_\gamma}{2},\quad
I_+=\frac{I_2-\imath I_1}{2}=\frac{\imath p_\gamma z+2\bar
\pi\left(1+z \bar z\right)}{4}e^{-2\imath \gamma},\quad
I_-=\overline{I}_+ \label{per}\ee

These quantities define, with respect to the Poisson brackets
(\ref{pb}), the $so(4)=so(3)\times so(3)$ algebra \be
\left\{P_i,P_j\right\}=\varepsilon_{ijk}P_k,\quad\left\{I_i,I_j\right\}=\varepsilon_{ijk}I_k,
\quad\left\{I_i,P_j\right\}=0. \label{pbso4}\ee The functions
$P_i$, $I_i$ obey the following equality, which is important for
our considerations \be I_i I_i=P_i P_i. \label{IP}\ee We
replace now the initial Lagrangian (\ref{lag0}) by the following
variationally equivalent one \be {\cal
L}_{int}=\left(P_+\Lambda_++P_-\Lambda_-+P_3\Lambda_3\right)+(\pi_\mu-P_i
A^i_\mu)\dot{x}_\mu-\frac{P_iP_i}{g r}-r \frac{ \pi_\mu
\pi_\mu}{g} +U(x) \label{lag7}.\ee
 Here we used the identity
$$-\frac{g r}{4}A_iA_i +g{\dot r}_\mu{\dot r}_\mu=
g\frac{\dot{x_\mu}\dot{x_\mu}}{4r}.$$ Such kind of representation
of a variationally equivalent Lagrangian is motivated by the
following reason. Let the 2n-dimensional Lagrangian have the form: \be
L=f_r(y)\dot{y}^r-H(y),\quad r=1\ldots 2n\label{llag}\ee where
$H(y)$ and $f_\mu(y)$ are some functions of the variable $y$. The
Euler-Lagrange equations for such Lagrangian look:
\be\dot{y}^r=\omega^{rs}\frac{\partial H}{\partial y^s},\ee where
$\omega^{rs}\omega_{sq}=\delta^r_q, \quad \omega_{rq}=\partial_r
f_q-\partial_q f_r$. It is easy to see that $\omega^{rq}$ defines
Poisson Brackets $\left\{y^r,y^s\right\}=\omega^{rs}$. Hence,
$\omega_{rq}$ is the symplectic structure of the system and $H$ is
the corresponding Hamiltonian.  In the invariant language we represent
the Lagrangian as follows: \be L=\omega^{(1)}(dy)-H(y),\ee where
$\omega^{(1)}$ is one-form which locally can be written as in
\ref{llag}, and the symplectic structure looks as follows:
\be\omega^{(2)}=d\omega^{(1)}.\ee We will use this fact in the
next section.

The isometries of this, new, Lagrangian corresponding to
(\ref{vec_2}) are defined by the vector fields \be {\bf\widetilde
U }_i\equiv \{I_i,\;\;\}, \label{is1}\ee where $I_i$ are given by
(\ref{per}) and the Poisson brackets are given by (\ref{pb}). Indeed,
$I_a$ are precisely the Noether constants of motion of the new
Lagrangian (\ref{lag7}) corresponding to (\ref{is1}). To see
this we simply should take into account the following equality
 \be
P_+\Lambda_++P_-\Lambda_-+P_3\Lambda_3=p_\gamma\dot{\gamma}+\pi\dot{z}+\bar\pi\dot{\bar
z}.\label{pp}\ee

Let us perform now the reduction by the action of the $SU(2)$ group given by the vector fields (\ref{is1}).
For this purpose we have to fix the Noether constants of motion(\ref{per}):
 $$I_i=s_i=const,\qquad s_is_i\equiv s^2 .$$
 Since  $I_i$ are constants of
motion independent on the $r_\mu$ coordinates we can perform an
orthogonal rotation,  so that only the third component of this set,
$I_3$,  will be different from zero. Equating $I_+$ and $I_-$ with
zero we obtain: \be -I_3=\frac{p_\gamma}{2}=s,\qquad \bar \pi=
-s\imath \frac{z}{1+z\bar z}, \quad \pi= s \imath \frac{\bar
z}{1+z\bar z}.\quad  \ee Hence,
 \be
P_+=-s\frac{\imath z}{1+z \bar{z}},\quad P_-=s\frac{\imath \bar
z}{1+z \bar{z}},\quad P_3=s\frac{1-z\bar z}{1+z\bar
z}\label{fix}\ee Thus $P_a$ become precisely the Killing
potentials of the $S^2$ sphere! It is not an occasional
coincidence, indeed.

Taking in mind the equality (\ref{pp}) we conclude that the
third term in (\ref{lag7}) can be ignored because it is a full time
derivative. Also, taking into account  \ref{IP} and denoting
${\tilde g}=g/2r$, one can rewrite the Lagrangian (variationally
equivalent) as follows:
 \be{\cal
L}^{red}_{int}=(\pi_\mu-sA^i_\mu h_i)\dot{x}_\mu-\imath
s\frac{\bar {z}\dot{z}-z \dot{\bar z}}{1+z\bar z}- \frac{ \pi_\mu
\pi_\mu}{2\tilde g} -\frac{s^2}{2{\tilde g}r}-U( {\bf x}),\qquad
\mu=1,\ldots,5.
\label{ll}\ee According to the definition, the reduced
Lagrangian can be obtained after performing a variation procedure
in terms of the variables $\pi_\mu$.
 The
second term in this reduced Lagrangian is the one-form defining
the symplectic (and K\"ahler) structure on $S^2$, in agreement with
the above observation that $P_a$ results in the Killing potentials
of $S^2$.

Thus, the Noether constants of motion do not allow us to exclude
the $z,\bar z$ variables. However, their time derivatives appear in the
Lagrangian in a linear way only, defining the internal degrees
of freedom of the five-dimensional isospin particle interacting
with the Yang monopole. As a consequence, the dimensionality of the
phase space of the reduced system is $2\cdot 5+2=12$. Only in
the particular case $s=0$, corresponding to the absence of Yang
monopole, we arrive at a five-dimensional system. Hence,  {\sl
locally},  the Lagrangian of the system can be formulated in the
six-dimensional space. We will use this fact in the next section.

\setcounter{equation}{0}
\section{Higgs  oscillator and (pseudo)spherical Yang-Coulomb system}

Let us apply the above construction to the Higgs oscillator on the eight-dimensional sphere and pseudosphere and
obtain, for the reduced system, the (pseudo)spherical generalization of the Yang-Coulomb system, in the spirit of \cite{pogos}.

For this purpose we introduce the conformal flat coordinates of $d-$dimensional
(pseudo)sphere, which are precisely the stereographic coordinates.
These coordinates are related with the Cartesian   coordinates of the ambient
$(d+1)$-dimensional space as follows (here and in the following we assume the unit radius of the sphere and pseudosphere)
\begin{equation}
 x_{ A}=\frac{2u_{  A}}{1+\epsilon  u^2_{ B} },\quad
 x_{d+1}=\frac{1-\epsilon u^2_B}{1+\epsilon u^2_B},\qquad{  A},{B}=1,\ldots d.
\label{x}\end{equation}
Here   ${ x}_A, x_{d+1}$ are the
(pseudo)Euclidean coordinates of the ambient space
$\DR^{d+1}$($\DR^{d.1}$): $\epsilon{ x}^2_A+ x^2_{d+1}=1$,
$\epsilon=\pm 1$.
 The $\epsilon=+1$ corresponds
to the sphere and  $\epsilon=-1$ corresponds to  the pseudosphere.

In these coordinates the metric takes the conformally-flat form  
\begin{equation}
 ds^2=\frac{4du_A du_A}{(1+\epsilon u^2_B)^2}, 
\label{met}\end{equation}
while
the potentials of the Higgs oscillator and of the
Schroedinger-Kepler system (\ref{v}) read
\begin{equation}
V_{osc}=\frac{2\omega^2 {\bar u}u}{(1-\epsilon {\bar u}u)^2},
\quad V_{C}= -\gamma\frac{1-\epsilon {\bar u}u}{2|u|}, \label{v1}
\end{equation}

 Hence, the $su(2)$-reduction of the the Higgs oscillator on the (pseudo)sphere to a $5d$ system
 leads to the following Lagrangian (compare with Ref. \ref{ll})\be
{\cal L}^{red}_{osc}={\cal L}_{red}=(\pi_\mu-sA^i_\mu
h_i)\dot{x}_\mu-\imath s\frac{\bar {z}\dot{z}-z \dot{\bar
z}}{1+z\bar z}-\frac{(1+\epsilon r)^2}{4} r\left(\frac{ \pi_\mu
\pi_\mu}{2} +\frac{s^2}{2r^2}\right) -\frac{2\omega^2
r}{(1-\epsilon r)^2} \label{losc}\ee
The one-form corresponding to
this Lagrangian have the following form:
\be \omega^{(1)}=\pi_\mu
dx_\mu-s A^i_\mu h_i dx_\mu +\imath s (h_-d z-h_+ d{\bar z}),\ee
and the inverse matrix of corresponding symplectic structure
$\omega^{(2)}=d\omega^{(1)} $ defines the  Poisson brackets
 \be
\left\{\pi_\mu,\pi_\nu\right\}=s\left(\partial_\mu A_\nu^i
h_i-\partial_\nu A_\mu^i h_i-\varepsilon_{ijk}h_i A^j_\mu
A^k_\nu\right)\equiv sF^i_{\mu\nu}h_i,\quad \left\{z,\bar
z\right\}=\frac{\imath}{2s}(1+z\bar z)^2,\quad
\left\{h_i,h_j\right\}=\frac{1}{s}\varepsilon_{ijk}h_k. \ee
The Hamiltonian of the reduced system is given by the expression
\be{\cal H}_{red}^{osc}=\frac{(1+\epsilon
r)^2}{4} r\left(\frac{ \pi_\mu \pi_\mu}{2}
+\frac{s^2}{2r^2}\right) +\frac{2\omega^2 r}{(1-\epsilon r)^2}.\ee

 On the other hand, the
Higgs oscillator has a number or symmetries: besides the
rotational $so(8)$ symmetries, defining the Noether constants of
motion, it possesses constants of motion which are quadratic in
momenta. We are interested in their $su(2)$ invariant subset given
by the generators \be J_{\mu\nu}=\frac{{\cal
P}\left[\Gamma_\mu,\Gamma_\nu\right]u}{2} \ee and \be {\bf
A}=\frac{J\Gamma_\mu J}{2} +2\frac{\omega^2 u\Gamma_\mu
u}{(1-\epsilon \bar u u)^2}\label{dt} \ee where $J_A=(1-\epsilon
u^2){\cal P}_A+2\epsilon ({\bf u}{\bf {\cal P}})u_A $ and ${\cal
P}_A$ is the corresponding momenta of the coordinate $u_A$.

Reducing the generators of rotations  to the $5d$ system,
following the general procedure described in the previous section,
we get \be J_{\mu\nu}=x_\mu \pi_\nu-x_\nu \pi_\mu+2r^2F_{\mu\nu}^i
h_i.\ee In order to find the expressions for the hidden symmetry
generators we exclude the subset of generators of rotations that
leaves invariant the coordinate $x_\mu$: \be {\cal
J}^\mu_{\alpha\beta}=\varepsilon_{\mu\alpha\beta\nu\lambda}J_{\nu\lambda}
,\quad \alpha,\beta=1\ldots 4,\quad \mu,\nu,\lambda=1\ldots 5
\label{pri} \ee
 Now, we can write the
implicit expression for $A_\mu$ in the following form: \be
A_\mu=\frac{J_{\mu\nu}T_\nu}{2}+\frac{q_\mu}{4}\left(1+\epsilon
r\right)^2\left(\pi_\mu\pi_\mu
+\frac{s^2}{r^2}\right)+\frac{2\omega^2  q_\mu}{(1-\epsilon
r)^2}+\frac{\epsilon}{2}\varepsilon_{\alpha\beta\gamma\delta}{\cal
J}^\mu_{\alpha\beta}{\cal J}^\mu_{\gamma\delta}, \ee where
$T_\mu=(1+q^2)p_\mu-2({\bf q}{\bf p})q_\mu$- denote transition
operators on the pseudosphere. Since the last terms is expressed
through the already mentioned motion integrals, we can ignore
them.

Following (\cite{pogos}), we can now transform the reduced Higgs
oscillator to a Kepler-like system. For this purpose we should
fix the energy surface ${\cal H}^{red}_{osc}=E\equiv  \gamma/2$
and multiply by $(1-\epsilon r)^2/r$, to get \be \left({\cal
H}^{red}_{osc}-E_{osc}^{red}\right)\frac{(1-\epsilon
r)^2}{r}=0\equiv{\cal H}_{MICZ}-E_{MICZ},\quad
E_{MICZ}=-\epsilon\gamma-2\omega^2\ee\be{\cal H}_{MICZ}=
\frac{\left(1- r^2\right)^2
}{4}\frac{1}{2}\left(\pi_\mu\pi_\mu+\frac{s^2}{
r^2}\right)+\gamma\frac{1+r^2}{2r} \ee For any motion integral
${\cal I}$ we have: \be\left\{H_{MICZ},{\cal
I}\right\}=\left\{H^{red}_{osc},{\cal
I}\right\}|_{H^{red}_{osc}=const}=0\ee  Hence, the new Hamiltonian
has the same number of motion integrals and therefore preserves
the integrability of the initial system. After fixing the energy surface
the quantities $A_\mu $ transform, respectively, in the Runge-Lenz
vector of the constructed $SU(2)$-Kepler (or Yang-Coulomb) system
on pseudosphere \be A_\mu=\frac{J_{\mu\nu}T_\nu}{2}+\gamma
\frac{q_\mu}{r}. \ee

Thus, we constructed the five-dimensional
pseudospherical generalization of the MICZ-Kepler system, i.e.
the pseudospherical Yang-Coulomb system, and found its constants of
motion. It is not difficult to find a spherical analog of this
system. Performing the Wick rotation, we obtain  the spherical
Yang-Coulomb system. It will be defined with the same Poisson
brackets as before by the Hamiltonian \be {\cal
H}_{MICZ}^{(sph)}=\frac{\left(1+ r^2\right)^2
}{4}\frac{1}{2}\left(\pi_\mu\pi_\mu+\frac{s^2}{
r^2}\right)+\gamma\frac{1-r^2}{2r}, \ee and by the motion
integrals $A_\mu$, where the quantities $T_\mu$ are replaced by
$T_\mu=(1-q^2)p_\mu+2({\bf q}{\bf p})q_\mu$
\subsection{Anysotropic inharmonic oscillator and Yang-Coulomb-Stark system}
The oscillator described in the previous section can be extended
by adding anisotropic and inharmonic parts \cite{ny}.

\be{\cal L}_{AIOSC}=\frac{4}{\left(1+\epsilon
u^2\right)^2}\frac{\dot{u}_i\dot{u}_i}{2}-\frac{2\omega^2
u^2}{(1-\epsilon u^2)^2}-\frac{2\Delta \omega^2
u\Gamma_5u}{(1-\epsilon
u^2)^2}-\frac{4\varepsilon_{el}}{(1-(u^2)^2)^2}\frac{1+(u^2)^2}{(1-\epsilon
u^2)^2}u^2 (u\Gamma_5 u)\ee Since the additional terms in this
Lagrangian do not preserve the spherical symmetry of the previous
system, only a part of the integrals of motion will be generalized. So, instead of $N=5(5-1)/2=10$ motion integrals
corresponding to the rotation symmetry $J_{\mu\nu}$, we have only
$N^\prime=4(4-1)/2=6$ ones,  $J^5_{\alpha\beta}$, defined in
\ref{pri}. Only one component of the generator of hidden symmetry
is generalized. In the Hamiltonian approach it has
the following form: \be A_5=\frac{J\Gamma_5 J}{2} +2\frac{\omega^2
u\Gamma_5 u}{(1-\epsilon \bar u u)^2}+2\frac{\Delta\omega^2
u^2}{(1+\epsilon \bar u u)^2}
+4\varepsilon_{el}\left(\frac{(u^2)^2}{(1-(u^2)^2)^2}+\frac{(u\Gamma_5u)^2}{(1-\epsilon
u^2)^4}\right)\ee It is obvious that the expression for this
quantity in the Lagrangian approach can be obtained just by replacing
the momenta by the corresponding time derivative divided by
$(1+\epsilon u^2)^2$. In the same way as we obtained the MICZ-Kepler system
in the previous section, we get now: \be
{\cal H}_{MICZ}=\frac{\left(1- r^2\right)^2
}{4}\frac{1}{2}\left(\pi_\mu\pi_\mu+\frac{s^2}{
r^2}\right)+\gamma\frac{1+r^2}{2r}+2\Delta\omega^2\left(\frac{1-\epsilon
r}{1+\epsilon r}\right)^2\frac{x_5}{r}
+2\varepsilon_{el}\frac{1+r^2}{1-r^2}\frac{x_5}{1-r^2}\label{MKS}\ee

\be
A_5=\frac{J_{5\nu}T_{\nu}}{2}+\gamma\frac{x_5}{r}+2\frac{\Delta\omega^2}{\left(1+\epsilon
r\right)^2}\frac{r^2-x_5^2}{r}+2
\varepsilon_{el}\frac{r^2-x_5^2}{\left(1-r^2\right)^2} \ee
 The fourth term in the Hamiltonian
\ref{MKS} in the limit of flat space yields a homogeneous Electric
field with strength $\varepsilon_{el}$. Hence, we can consider it
as a generalization of the Stark term in case of a
(pseudo)-spherical space. The third term is just the
$\cos{\theta}$ potential.

As in the previous section, the transition to the sphere can be realized
by performing a Wick rotation. All terms in \ref{MKS} result in real
expressions,except the third one. However, one can notice that we
can consider the real and imaginary parts of the Hamitonian
separately. Indeed, let the Hamiltonian ${\cal H}$ and a motion
integral ${\cal I}$ have the following form:
$${\cal H}=K_H+U+{\bf i}V,\quad {\cal I}=K_I+P+{\bf i}Q,$$
where $U,V,P,Q$ are real functions of  coordinates and $K_H$ and
$K_I$ are kinetic terms of Hamiltonian and motion integral
respectively. The condition $\left\{{\cal H},{\cal I}\right\}=0$
leads us to two equalities: $$\left\{{\cal H}_{Re},{\cal
I}_{Re}\right\}=0,\quad\left\{{\cal H}_{Im},{\cal
I}_{Im}\right\}=0,
$$
where ${\cal H}_{Re}=K_H+U,\quad {\cal I}_{Re}=K_I+P$ and ${\cal
H}_{Im}=K_H+V,\quad {\cal I}_{Im}=K_I+Q.$ This still will not lead
us to a new system. After separating the expression we will find
that its imaginary part looks exactly like the Stark term and
therefore can be ignored. Explicitly we find: \be {\cal
H}_{MICZ}^{sph}=\frac{\left(1+ r^2\right)^2
}{4}\frac{1}{2}\left(\pi_\mu\pi_\mu+\frac{s^2}{
r^2}\right)+\gamma\frac{1+r^2}{2r}+2
Re\Delta\omega^2\left(\frac{1-\imath\epsilon r}{1+\imath\epsilon
r}\right)^2\frac{x_5}{r}
+2\varepsilon_{el}\frac{1-r^2}{1+r^2}\frac{x_5}{1-r^2}\label{MKS1}\ee
\section{Conclusion and discussion}
In this paper we applied the $su(2)$ reduction procedure to the Higgs oscillator on the eight-dimensional pseudosphere and sphere and,
transforming the energy level of the reduced system, we obtained the Yang-Coulomb systems on the five-dimensional
 sphere and pseudosphere. We remind that the Yang-Coulomb system is the generalization of the five-dimensional Coulomb
 system specified by the
  presence of $SU(2)$ Yang monopole, which inherits the symmetries of the five-dimensional Coulomb system.
Similarly, the constructed (pseudo)spherical Yang-Coulomb system inherits all the symmetries of the five-dimensional
(pseudo)spherical Coulomb system.
We also applied this procedure to the anisotropic inharmonic Higgs oscillator \cite{anosc} and obtained, in this way, the integrable
(pseudo)spherical generalization of the Yang-Coulomb system with Stark term. While the spectrum of the previous system can be easily
 obtained from the general construction \cite{mny}, the spectrum (even perturbative)
  of the latter one needs to be constructed. Even for the planar case it is unknown up to now. We are planning to
investigate it in forthcoming studies.

There are  related problems that definitely need to be studied.
Few years ago the analog of oscillator on complex projective space $CP^N$ has been suggested, which respected the inclusion of
constant magnetic field \cite{cpn}. It was found, that in the  $CP^2$ case this oscillator can be reduced to the (three-dimensional)
MICZ-Kepler system on (pseudo)sphere. It is interesting to clarify whether the oscillator on $C P^4$ results, upon
$su(2)$ reduction, to the (pseudo)spherical Yang-Coulomb system. Also, in complete similarity to $CP^N$,
one can define the oscillator
on the quaternionic projective space $HP^N$ respecting the inclusion of instanton field \cite{hpn}.
One can expect that the $su(2)$reduction of such an oscillator on $HP^2$ would also yield a (pseudo)spherical
 Yang-Coulomb system. Both these statements should be carefully checked.

{\bf \large Acknowledgments.} The authors thank Armen Nersessian for the collaboration at the initial stage of the work,
 careful reading of manuscript, numerous discussions and comments.
V.~Y. acknowledges LNF for the hospitality and financial
support during his stay in Frascati, where the part of the present work was carry out.
The work was  partially supported by the
NFSAT-CRDF UC 06/07 (V.~Y.), ANSEF PS1730 ( V.~Y.) and  INTAS-05-7928 (S.~B, V.~Y.) grants.


\begin{thebibliography}{99}
\bibitem{ta}V.~M.~Ter-Antonyan, { Dyon-Oscillator duality},
{\it Lectures given at  BLTP International School on Symmetries and Integrable Sysems, Dubna, Russia, 8-11 Jun 1999}
[quant-ph/0003106]

 \bibitem{bohlin}
T. Levi-Civita, Opere Mathematiche, {\bf 2}, 411 (1906);

K.~Bohlin, Bull.~Astr., {\bf 28}  144 (1911)

 \bibitem{ks}P.~Kustaanheimo,~E.~Stiefel,~J.~Reine Angew Math.,
 {\bf 218}, 204 (1965);
 \bibitem{h}A.~Hurwitz,~Mathematische~Werke,~Band~II,~641 ({\sl Birkh\"auser},~Basel,~1933)

L.~S.~Davtyan {\it et al}, J.~Phys.~{\bf A20}~6121~(1987)
\bibitem{ntt}A.~Nersessian, V.~Ter-Antonian,  M.~M.~Tsulaia, Mod.\ Phys.\ Lett.\  {\bf A11} (1996), 1605
\bibitem{micz} D. Zwanziger, {Phys. Rev.\/} {\bf 176}, 1480
(1968)

H. McIntosh and A. Cisneros, {J. Math. Phys.\/} {\bf 11}, 896
(1970).


\bibitem{su2}T.~Iwai,
 J.\ Geom.\ Phys. {\bf 7}, 507 (1990);
L.~G.~Mardoyan,
 A.~N.~Sissakian, V.~M.~Ter-Antonyan, Phys.\ Atom.\ Nucl.
 {\bf 61}, 1746 (1998)


\bibitem{sphere}E.~Schr\"odinger,~Proc.~Roy.~Irish~Soc.~{\bf 46}~9~(1941); {\bf 46}~183
(1941); {\bf 47}~53~(1941)
\bibitem{sphere1}P.~W.~Higgs,~J.~Phys.~A:~Math.~Gen.~{\bf 12}~309~(1979)

H.~I.~Leemon,~J.~Phys.~A:~Math.~Gen.~{\bf 12}~489~(1979)
\bibitem{garb}A.~Barut, A.~Inomata, G.~Junker,~J.~Phys. {\bf A20},~6271~(1987);
J.~Phys.~{\bf A23}~1179~(1990),

Ya.~A.~Granovsky,~A~.S.~Zhedanov,~I.~M.~Lutzenko,
Teor.~Mat.~Fiz.{\bf 91}~207~(1992); {\bf 91}~396~(1992);

D.~Bonatos,~C.~Daskaloyanis, K.~Kokkatos, Phys.~Rev. {\bf A50}~3700~(1994);

E.~G.~Kalnins,~W.~Miller~Jr.,~G.~S.~Pogosyan, J.~Math.~Phys.
{\bf 37}~6439~(1996);{\bf 38}~5416~(1997)

 A. Ballesteros, F. J. Herranz, O. Ragnisco,  J.Phys. {\bf A38}  7129 (2005);

A. Ballesteros, F. J. Herranz, J.Phys. {\bf A40} F51(2007)


J. F. Carinena, M. F. Ranada, M. Santander, Ann. Phys. {\bf 322},  2249 (2007)

\bibitem{pogos} A.~Nersessian and G.~Pogosyan,
 { Phys.\ Rev.\/} {\bf A63}, 020103(R) (2001)

 A.~Nersessian,
{Phys.\ Atom.\ Nucl.\/}  {\bf 65}, 1070 (2002) ;

 \bibitem{kurochkin}
V.~V.~Gritsev, Yu.~A.~Kurochkin, V.~S.~Otchik,
{ J.~Phys.\/} {\bf A33}, 4903-4910 (2000)

\bibitem{anosc}
 A.~Nersessian and V.~Yeghikyan,
  J.\ Phys.\ {\bf A41}, 155203 (2008)


  \bibitem{2nd}  M.~Gonzales, Z.~Kuznetsova, A.~Nersessian, F.~Toppan and V.~Yeghikyan,
  {\sl ``Second Hopf map and supersymmetric mechanics with Yang monopole,''}
  arXiv:0902.2682 [hep-th].




\bibitem{mny}
  L.~Mardoyan, A.~Nersessian and A.~Yeranyan,
  Phys.\ Lett.\  A {\bf 366}, 30 (2007)
  \bibitem{stark}  L.~Mardoyan {\sl et al} 
  Theor.\ Math.\ Phys.,  {\bf 140},  958 (2004);
         { J.\ Phys.} {\bf A40}, 5973 (2007)  

S.~Bellucci and V.~Ohanyan, 
  Phys.\ Lett.\  A {\bf 372}, 5765 (2008)
\bibitem{cpn} S.~Bellucci and A.~Nersessian,
  Phys.\ Rev.\   {\bf D67} (2003) 065013
  [Erratum-ibid.\  {\bf D71} (2005) 089901];

S.~Bellucci, A.~Nersessian and A.~Yeranyan,
  Phys.\ Rev.\  {\bf D70} (2004) 045006;

{\sl ibid.} 
 {\bf 70} (2004) 085013

    {\sl ibid.} 
 {\bf 74} (2006) 065022;

  S.~Bellucci and A.~Nersessian,
  {\sl ``Supersymmetric Kahler oscillator in a constant magnetic
  field,''},
 {\it Contributed to International Seminar on Supersymmetries and Quantum
 Symmetries SQS 03, Dubna, Russia, 24-29 Jul 2003},
  hep-th/0401232;

S.~Bellucci and A.~Nersessian, Nucl. Phys. Proc. Suppl. {\bf 102}
(2001) 227; S.~Bellucci and A.~Nersessian,
 Phys.\ Rev.\  {\bf D64} (2001) 021702.
  \bibitem{hpn}L.~Mardoyan and A.~Nersessian,
  Phys.\ Rev.\ {\bf B72} (2005) 233303

 S.~Bellucci, P.-Y.~Casteill and A.~Nersessian,
  Phys.\ Lett.\  {\bf B574} (2003) 121

 S.~Bellucci, L.~Mardoyan and A.~Nersessian,
  Phys.\ Lett.\  {\bf B636} (2006) 137



\end{thebibliography}
\end{document}